\documentclass{PoS}

\usepackage{graphicx}
\usepackage{amsmath,epsfig}
\usepackage{amssymb,amsfonts}
\usepackage{latexsym}

\relax

\def\be{\begin{equation}}
\def\ee{\end{equation}}

\newcommand{\la}{\lambda}

\newcommand{\bear}{\begin{eqnarray}}
\newcommand{\bea}{\begin{eqnarray}}
\newcommand{\eear}{\end{eqnarray}}
\newcommand{\eea}{\end{eqnarray}}

\def\hri#1#2{\href{http://arxiv.org/abs/#1}{[ArXiv:#1]#2}}
\def\hre#1#2{\href{http://arxiv.org/abs/#1/#2}{[ArXiv:#1/#2]}}

\newbox\pippobox

\def\II{\relax{\rm I\kern-.18em I}}

\def\l{\lambda}
\def\m{\mu}
\def\n{\nu}

\def\h{\kappa}
\def\gf{w}

\title{The spectrum of (h)QCD in the Veneziano limit}

\ShortTitle{Spectrum of V-QCD}

\author{\speaker{Daniel Are\'an}
\\
\href{http://www.ictp.it/}{International Centre for Theoretical Physics (ICTP)}
and INFN - Sezione di Trieste \\
Strada Costiera 11; I 34014 Trieste, Italy\\
        E-mail: \email{arean@sissa.it}}

\author{Ioannis Iatrakis\\
        \href{http://hep.physics.uoc.gr}{Crete Center for Theoretical Physics},
Department of Physics, University of Crete, 71003 Heraklion, Greece\\
        }
\author{Matti J\"arvinen\\
        \href{http://hep.physics.uoc.gr}{Crete Center for Theoretical Physics},
Department of Physics, University of Crete, 71003 Heraklion, Greece\\
        }

\abstract{

In this note we report on 
the analysis of 
the zero temperature spectra of glueballs and mesons for holographic QCD in the Veneziano limit. 
We work within a  holographic bottom-up model named V-QCD which takes into account the full backreaction of the
flavor degrees of freedom. By studying the fluctuations of this model we compute spectra of mesons and
glueballs as a function of $x={N_f / N_c}$.
The spectra are discrete and gapped (modulo the pions) in the QCD regime, where $x$ is below the critical
value $x_c$ at which the conformal transition takes place.
The masses uniformly converge to zero in the walking region $x\to x_c$ following Miransky scaling.
Moreover, all the ratios of masses asymptote to finite constants as $x\to x_c$. Therefore there is no
``dilaton'' in the spectrum. Finally, we compute the S-parameter, which is found to be of $\mathcal{O}(1)$ in the
walking regime.}

\FullConference{CCTP-2013-03\vspace{.15cm}
\\Proceedings of the Corfu Summer Institute 2012 ``School and Workshops on Elementary Particle Physics and Gravity''\\
		September 8-27, 2012\\
		Corfu, Greece}

\begin{document}

\section{Introduction}
The gauge/gravity duality \cite{gg} is a powerful tool to study the dynamics of strongly coupled field theories and thus it has
been extensively used to try and describe the physics of Strong interactions. Although a calculable
string dual of QCD is far from our present understanding, it is possible to build models describing strong coupling
physics bearing many similarities with QCD. Indeed, top-down models resulting from solutions of string theory
describe qualitatively the low energy dynamics of QCD. However, these models are plagued by Kaluza-Klein modes
that make difficult a quantitative matching to real QCD.
A different approach is given by the bottom-up models, which are ad hoc holographic models inspired by string
theory that use some QCD features as inputs.
(See \cite{dis} and references therein for an overview on holographic duals of QCD).
In this note we will be working with a class of bottom-up holographic theories  that has been proposed
recently, under the name of V-QCD \cite{jk}, and has physics very close to  QCD in the Veneziano
limit.

The Veneziano limit \cite{veneziano} of QCD is given by:
\be
N_c\to\infty\,,\qquad N_f\to\infty\,,\qquad x={N_f\over N_c}\quad {\rm fixed}\,,\qquad
\lambda=g_{YM}^2\,N_c \quad {\rm fixed}\,.
\ee
An interesting feature of the theory accessible in this limit is the conformal window in which the theory has
an IR fixed point. This window extends from $x=11/2$ to lower values of $x$, and it includes the Banks-Zaks
weakly coupled region (as $x\to11/2$) \cite{bz}. At a critical value $x_c$ there is a phase transition from
the conformal window to theories with chiral symmetry breaking in the IR. Interestingly, near and below
$x_c$ there is a transition region where the theory is expected to exhibit ``walking'' behavior. This ``walking'' regime has been conjectured to display Miransky scaling \cite{miransky}.

The dynamics of ``walking'' (or nearly conformal) quantum field theories, has been the subject of intensive
study. It has been argued to be an important ingredient \cite{walk1,walk2,walk3} in providing
viable non-perturbative mechanisms for electroweak symmetry breaking like technicolor \cite{tech}.
As we have said, this regime is expected to appear in standard QCD just below the boundary of the
conformal window, $x\leq x_c\simeq 4$; as well as in other quantum field theories \cite{neil}.

The transition between the conformal window and QCD-like IR behavior has been called a conformal
transition \cite{conf}. It has been suggested that in holographic theories this conformal transition is
associated with a violation of the BF bound in the dual bulk theory \cite{son}.
Moreover, in QCD this correlates with the $\bar \psi \psi$ operators reaching a scaling dimension equal
to two -- another prerequisite of viable extended technicolor.

Apart from Miransky scaling, other phenomena have often been associated with the ``walking regime''
of QFT:
\begin{itemize}
 \item The appearance of an anomalously light scalar state, the ``dilaton'', due to the almost unbroken scale
 invariance \cite{walk2}.
 \item The suppression of the electroweak S-parameter, a crucial ingredient for the viability of
 technicolor theories \cite{as}.
\end{itemize}
Both issues are controversial, especially since  ``walking regimes'' appear at strong coupling, and therefore
perturbative techniques do not apply. It is also difficult to study these phenomena on the lattice due to
sizable finite size effects.

In recently studied holographic models with walking behavior, the lightest state is often a scalar
\cite{dilaton,kutasov}. Whether this state can be identified as the dilaton is, however, a difficult question,
and the answer appears to depend on the model. The S-parameter has been studied in popular
holographic bottom-up \cite{hong} as well as brane-antibrane models \cite{cobi,parnachev} with a
variety of answers found.

In this letter, to look into these and related issues we will study the spectra of mesons and glueballs for
a bottom-up holographic model of QCD in the Veneziano limit (V-QCD \cite{jk}).
This letter summarizes the results presented by one of the authors in the
``XVIII European Workshop on String Theory'', and subsequently published in \cite{letter}.

\section{V-QCD}

This model combines two sectors resulting from holographic models describing respectively the glue
and flavor dynamics of QCD.
 The first one is improved holographic QCD (IHQCD), which is a
  holographic model for large-N Yang Mills in 4 dimensions \cite{ihqcd}.
 The second one is a model for flavor inspired by tachyon condensation in string theory \cite{ckp}.
 The relevant fields that are kept in these models
 are: 
\begin{itemize}
 \item The five-dimensional metric, the space-time components of which are dual to the energy-momentum tensor,
 \item A scalar (the dilaton, $\phi$) that is dual
to the YM  't Hooft coupling constant, 
\item A complex $N_f\times N_f$ matrix field
(the tachyon, $T_{ij}$) transforming in the $(N_f,\bar N_f)$ of the
 $U(N_f)_L\times U(N_f)_R$ flavor group, and dual to the operator  $\bar \psi_j \psi_i$.
 \item Two gauge fields $A_M^{L}$ and $A_M^{R}$ transforming in the adjoint representation of $U(N_f)_L$ and
 $U(N_f)_R$ respectively. They are  $N_f\times N_f$ matrices dual to the $U(N_f)_{L,\,R}$ flavor currents, and thus
 they vanish for the vacuum solutions.
\end{itemize}


The complete action for the V-QCD model can be written as 
\be
 S = S_g + S_f + S_a\,,
\ee
where $S_g$, $S_f$, and $S_a$ are the actions for the glue, flavor and CP-odd sectors, respectively.
As discussed in \cite{jk}, only the first two terms contribute to the vacuum structure of the 
theory. The CP-odd sector, whose physics contains the $U(1)_A$ anomaly, contributes to some sectors
of the spectrum (flavor singlet pseudoscalars) that are not the subject of this letter; hence we will
address its physics in a future publication  \cite{to}. The full structure of the flavor sector action ($S_f$)
was not detailed in \cite{jk} since it is not necessary when studying the vacuum structure of the model.
However, the extra terms do contribute to the spectrum of fluctuations, and will be discussed below.

\subsection{The glue sector} \label{sec:VQCDglue}

The glue action was introduced in \cite{ihqcd,ihqcd2,data},
\be
S_g= M^3 N_c^2 \int d^5x \ \sqrt{-g}\left(R-{4\over3}{
(\partial\lambda)^2\over\lambda^2}+V_g(\lambda)\right) \, .
\ee
Here $\l=e^\phi$ is the exponential of the dilaton. It is dual to the ${\mathbb Tr} F^2$  operator,
and its background value is identified as the 't Hooft coupling.
Glue dynamics sets requirements on the dilaton potential: $V_g$ asymptotes to a constant near
$\la=0$, and diverges as $V_g\sim \l^{4 / 3}\sqrt{\log\la}$ as $\l\to \infty$, generating confinement,
a mass gap, discrete spectrum and asymptotically linear glueball trajectories \cite{ihqcd,data}. The Ansatz
for the vacuum solution for the metric is
\be
ds^2=e^{2 A(r)} (dx_{1,3}^2+dr^2)\,,
\label{bame2}
\ee
where the warp factor $A$ is identified as the logarithm of the energy scale in field theory. In our
conventions the UV boundary lies at $r=0$, and the bulk coordinate runs from zero to infinity. The
metric will be close to the AdS one except near the IR singularity at $r=\infty$, and thus
$A \sim -\log(r/\ell)$, where $\ell$ is the AdS radius. Therefore, $r$ is identified roughly as the inverse
of the energy scale of the dual field theory.

\subsection{The flavor sector} \label{sec:VQCDflavor}

The flavor sector consists of the generalized Sen's action \cite{senreview}
\be
S_f= - \frac{1}{2}\, M^3\, N_c\, {\mathbb Tr} \int d^4x\, dr\, \left( V_f(\l,T^\dagger T)
\sqrt{-\det {\bf A}_L}+ V_f(\l,TT^\dagger)\sqrt{-\det {\bf A}_R}\right)\,.
\label{generalact}
\ee
The quantities inside the square roots are defined as
\begin{align}
{\bf A}_{L\,MN} &=g_{MN} + \gf(\l) F^{(L)}_{MN}
+ {\h(\l) \over 2 } \left[(D_M T)^\dagger (D_N T)+
(D_N T)^\dagger (D_M T)\right] \,,\nonumber\\
{\bf A}_{R\,MN} &=g_{MN} + \gf(\l) F^{(R)}_{MN}
+ {\h(\l) \over 2 } \left[(D_M T) (D_N T)^\dagger+
(D_N T) (D_M T)^\dagger\right] \,,
\label{Senaction}
\end{align}
where the fields  $A_{L}$, $A_{R}$, and $T$ are $N_f \times N_f$ matrices in the flavor space.
It is not known in general how the determinants over the Lorentz indices in~\eqref{generalact} should
be defined when the arguments~\eqref{Senaction} contain non-Abelian matrices in flavor space.
However, for our purposes such definition is not required: our background solution will be proportional
to the unit matrix $ \mathbf{1}_{N_f}$, in which case the fluctuations of the Lagrangian are unambiguous
up to quadratic order.
The covariant derivative of the tachyon field is defined as
\be
D_M T = \partial_M T + i  T A_M^L- i A_M^R T\,.
\ee
And the class of tachyon potentials that we will consider is
\be
V_f(\l,TT^\dagger)=V_{f0}(\l) e^{- a(\l) \,T T^\dagger} \, .
\label{tachpot}
\ee
For the vacuum solutions (with flavor independent quark mass) we will have
$T =\tau(r) \mathbf{1}_{N_f}$ where $\tau(r)$ is real, so that $V_f(\l,TT^\dagger)$ is replaced by
\be
 V_f(\l,\tau)=V_{f0}(\l) e^{- a(\l)\tau^2} \, .
\ee
The other undetermined functions in the flavor action 
must satisfy the requirements that we discuss below.

\subsection{Background solutions and phase diagram}
Agreement with the dynamics of QCD both in the IR and UV sets requirements on the undetermined
functions appearing in the action of V-QCD. In the UV, agreement with the two-loop QCD beta-function
and the one-loop anomalous dimension of the quark mass restricts the asymptotics of these potentials,
\cite{jk}.
Moreover, $V_g(\l)$ has already been fixed from glue dynamics \cite{data}, and the other
undetermined functions in the flavor action ($V_{f0}(\l)$, $\kappa(\l)$, $a(\l)$, $w(\la)$)
must satisfy the following generic requirements:
\begin{itemize}
\item There should be two extrema in the potential for  $\tau$: an unstable maximum at $\tau=0$
(with chiral symmetry intact) and a minimum at $\tau=\infty$ (with chiral symmetry broken).
\item The dilaton potential at $\tau=0$, namely $V_\mathrm{eff}(\la)=V_g(\la)-x\,V_{f0}(\la)$, must
have a nontrivial IR extremum at $\la=\la_*(x)$ that moves from $\la_*=0$
at $x={11/ 2}$ to large values as $x$ is lowered.
\end{itemize}

Notice that the Ansatz (\ref{tachpot}) for $V_f(\l,\tau)$ automatically satisfies the first requirement. On the other
hand, the second requirement is necessary for the phase diagram to have the required structure as a function
of $x=N_f/N_c$.
Different classes of potentials have been studied in \cite{jk,alho} where they were classified according
to the IR behavior of the tachyon as type I and type II models. These are the potentials that will be
used in the analysis presented in this letter. The potentials I and II in \cite{alho} are completely determined
up to a constant called $W_0$ which controls the flavor dependence of the UV AdS scale.
We refer to \cite{alho} for the explicit form and a detailed explanation of the potentials used in this
analysis\footnote{A
thorough study of the potentials, taking into account the constraints arising from the meson spectra,
will be performed in \cite{to}.}.

The vacuum (zero temperature) solutions of \cite{jk}
involve a Poin\-car\'e invariant metric, no vectors, and radially dependent scalars:
\be
ds^2=e^{2 A(r)} (dx_{1,3}^2+dr^2)\,,\quad \lambda(r)\,,\quad T=\tau(r)~\mathbf{1}_{N_f}\;.
\label{bame}
\ee
The background solutions corresponding to the aforementioned potentials result in a zero temperature phase
diagram that is essentially universal. As a function of $0<x<{11/ 2}$, the standard phase diagram at zero quark mass
presents two phases separated by a phase transition at some $x=x_c\simeq 4$ (the precise value depends on the
potential chosen):
\begin{itemize}
 \item In the region
 $0<x<x_c$, the (massless) theory has chiral symmetry breaking, and flows to a massless $SU(N_f)$ pion theory
 in the IR. The IR dynamics is thus similar to the one of ordinary QCD. The corresponding background solution
 has nontrivial $\l(r)$, $A(r)$ and $\tau(r)$, with the tachyon  diverging at the IR singularity of the geometry.
 
 \item In the conformal window, i.e., when $x_c<x<{11/2}$, the theory flows to a nontrivial IR fixed point and there
 is no chiral symmetry breaking.
  The background solution has zero tachyon $\tau(r)$=0 and nontrivial $\l(r)$ and $A(r)$, giving rise to a
 geometry flowing to a nontrivial AdS fixed point in the IR.
\end{itemize}

\begin{figure}[!tb]
\centering
\includegraphics[width=0.45\textwidth]{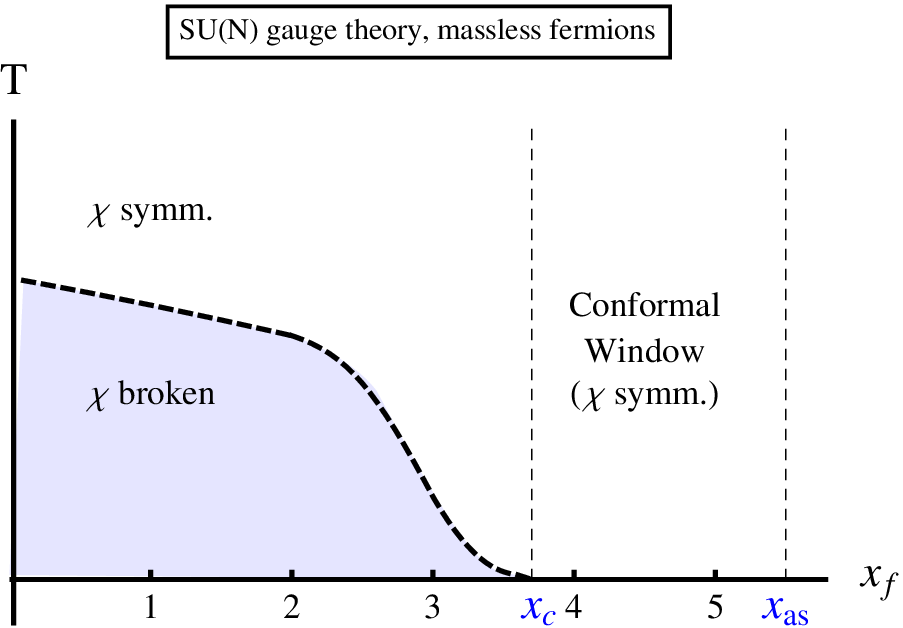}
\caption{\small Qualitative behavior of the transition temperature between the low and high
temperature phases of V-QCD matter \cite{alho}.
 }
\label{f1}
\end{figure}

Remarkably, in the region just below the conformal window ($x\lesssim x_c$) the theory exhibits
a ``walking'' behavior. The phase transition at $x=x_c$ (which is only present at zero quark mass)
involves BKT \cite{bkt} or Miransky \cite{miransky} scaling. Indeed, the order parameter of the transition,
the chiral condensate $\sigma \sim \langle \bar q q \rangle$, vanishes exponentially:
\be \label{condscaling}
 \sigma \sim \exp\left(-\frac{2 \hat K}{\sqrt{x_c-x}}\right)\,,
\ee
as $x \to x_c$ from below. (The constant $\hat K$ is positive \cite{jk}).
And for $x_c \ge x$, $\sigma$ is identically zero as chiral symmetry is intact.
The Miransky scaling is linked to the ``walking'' behavior of the coupling constant: the field $\l(r)$ takes an
approximately constant value $\l_*$ for a wide range of $r$  (the length of this region scales as the square
root of the condensate in~\eqref{condscaling}).
Hence, the coupling stays approximately constant for many decades in the RG time, until the deep IR
where the non-zero tachyon drives the theory away from the nontrivial fixed point and towards $\la=\infty$.

At finite temperature a rich structure of black holes, with one or two scalar hairs, was found in \cite{alho}.
The general structure of their phase diagram is depicted in figure \ref{f1}.
The chiral restoration transition is first order at low values of $x$, but typically becomes second order as
we approach $x_c$. When this happens, a separate first order deconfinement transition still exists,  so that
an extra chirally broken phase appears for temperatures between the two transitions. 
There can also be two first-order transitions, depending on the details of the potentials. The transition
temperatures obey Miransky scaling \cite{alho}.

\section{Quadratic fluctuations and spectra}

In this section we present some results for the spectra of mesons and glueballs of V-QCD, restricting to the case
of zero quark mass.


In order to compute the spectrum of mesons and glueballs we need to study the fluctuations of all the fields of
V-QCD. In the glue sector the relevant fields are the metric $g_{mn}$, the dilaton $\phi$ and the QCD axion
$a$ (which belongs to the CP-odd sector). Their normalizable fluctuations correspond to glueballs
with $J^{PC}=0^{++},\; 0^{-+},\; 2^{++}$, where $J$ stands for the spin and $P$ and $C$ for the field
properties under parity and charge conjugation respectively. In the meson sector one has the tachyon $T$,
and the gauge fields $A_\m^{L/R}$; their normalizable fluctuations corresponding to mesons with
$J^{PC}=1^{++},\; 1^{--},\; 0^{++},\;  0^{-+}$. 

\begin{figure}[t]
\begin{center}
\includegraphics[width=0.49\textwidth]{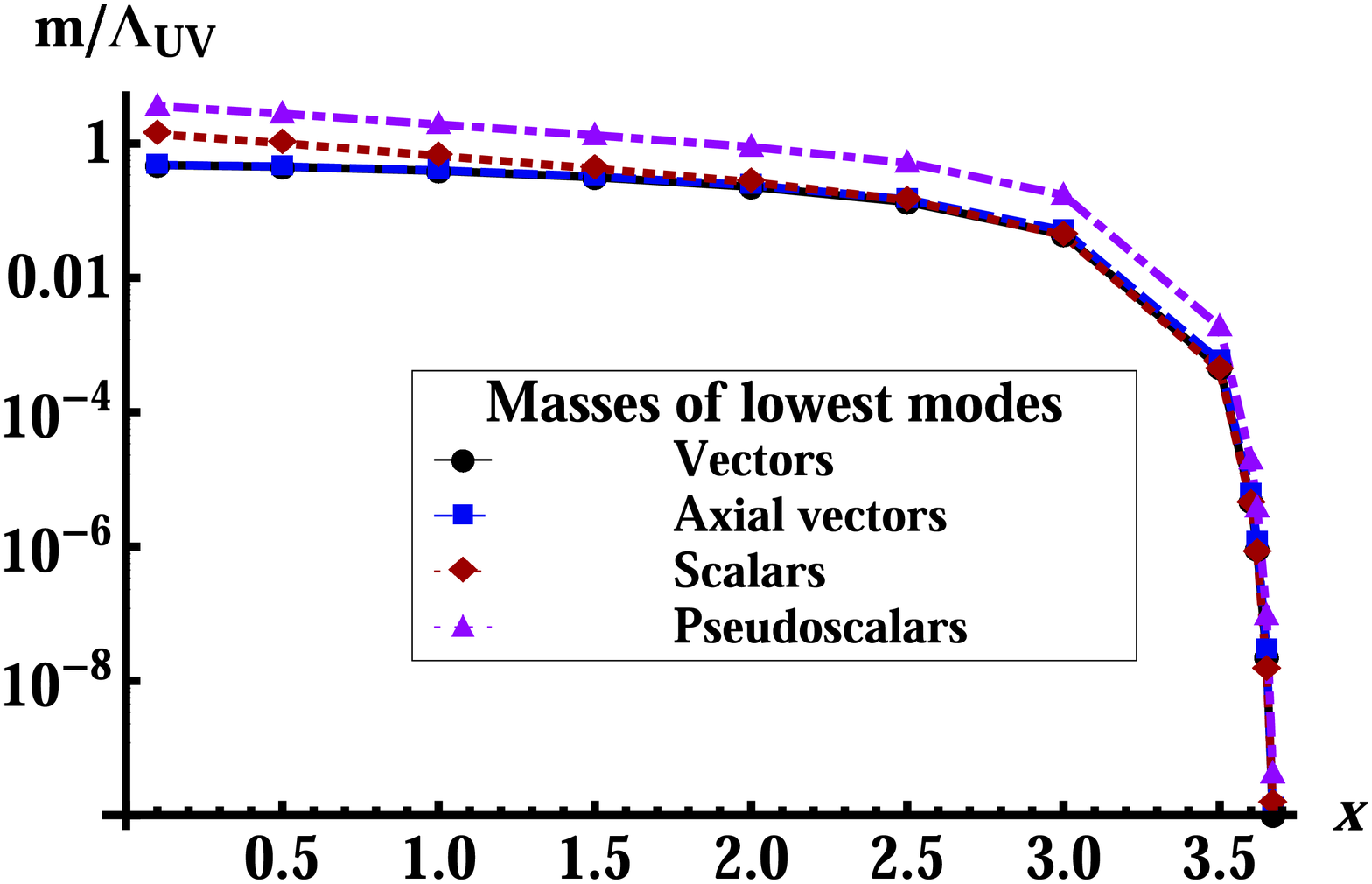}\hfill
\includegraphics[width=0.49\textwidth]{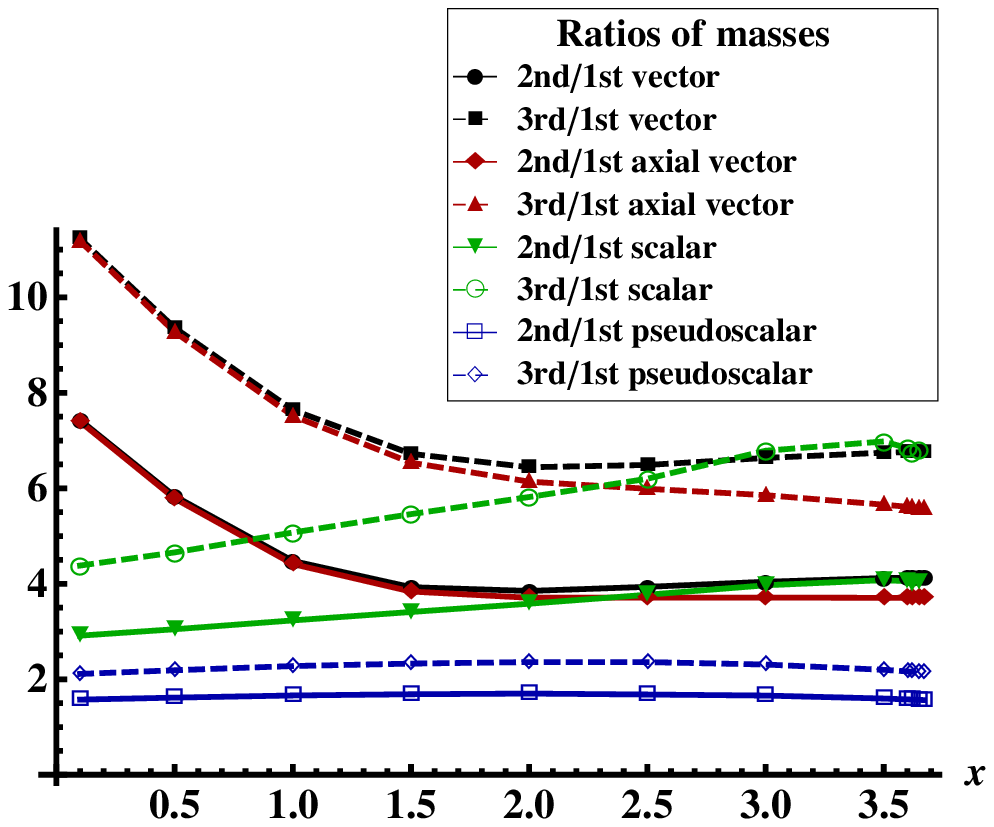}
\end{center}
\caption{\small Nonsinglet meson spectra in the potential II class with Stefan-Boltzmann (SB) normalization
for $W_0$ (see \cite{alho}), with $x_c \simeq 3.7001$. Left: the lowest non-zero masses of all four towers
of mesons, as a function of $x$, in units of $\Lambda_{\rm UV}$, below the conformal window. Right, the ratios
of masses of up to the fourth massive states in the same theory as a function of $x$.}
\label{f2}\end{figure}

\begin{figure}[t]
\begin{center}
\includegraphics[width=0.49\textwidth]{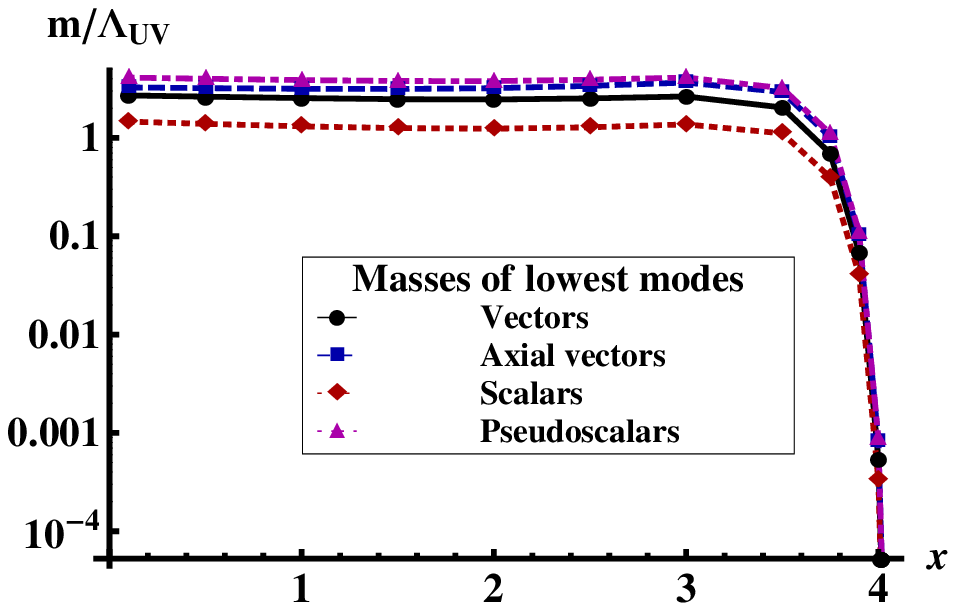}\hfill
\includegraphics[width=0.49\textwidth]{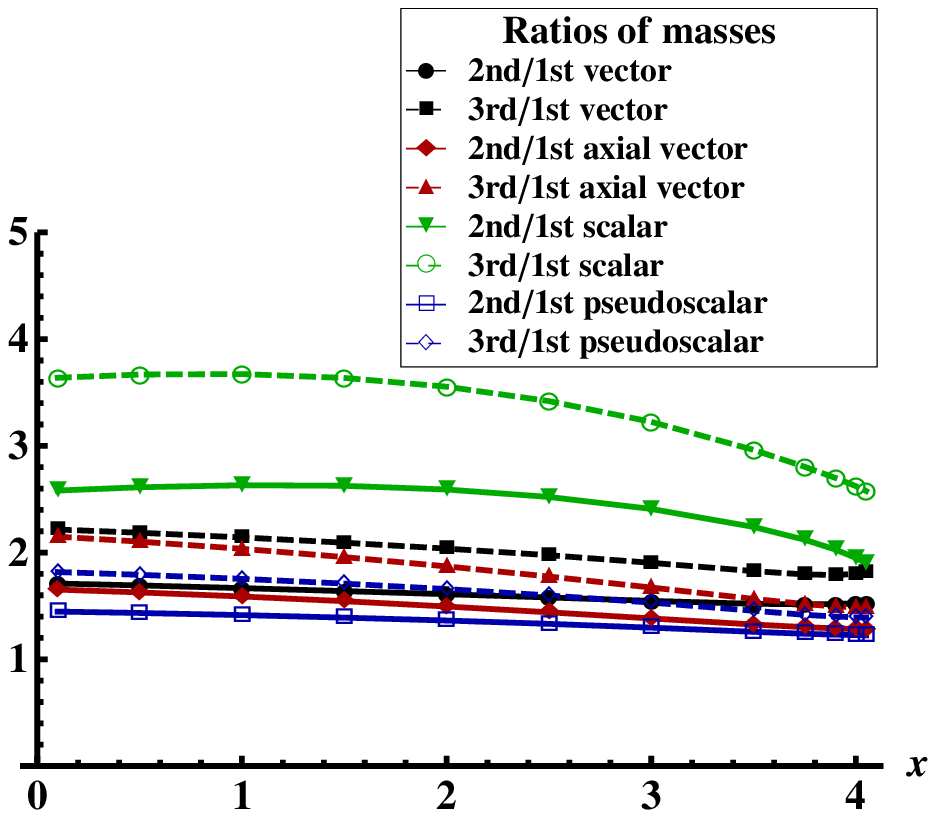}
\end{center}
\caption{\small Nonsinglet meson spectra in the potential I class ($W_0={3/ 11}$), with $x_c \simeq 4.0830$.
Left: the lowest non-zero masses of all four towers of mesons, as a function of $x$,
in units of $\Lambda_{\rm UV}$, below the conformal window. Right, the ratios of masses of up to the fourth
massive states in the same theory as a function of $x$.}
\label{f3}\end{figure}
The fluctuations fall into two classes according to their transformation properties under the flavor group: flavor
nonsinglet (transforming in the adjoint representation of $SU(N_f)$) modes and flavor singlet modes. The glue
sector contains only flavor singlet modes, whereas each fluctuation in the meson sector can be divided into
flavor singlet and nonsinglet terms. Those (flavor singlet) modes which are present in both sectors will mix.
Since we are in the Veneziano limit the mixing takes place at leading order in $1/N_c$: the $0^{++}$
glueball mixes with the $0^{++}$ flavor singlet $\sigma$-meson, and the pseudoscalar $0^{-+}$ flavor
singlet meson mixes with the $0^{-+}$ glueball due to the axial anomaly (realized by the CP-odd sector
which we will not study here, see \cite{to}). 
All classes, with various $J^{PC}$ and tranformation properties under the $U(N_f)$ group, contain an infinite
discrete tower of excited states.

We  start by defining the vector and axial vector combinations of the gauge fields:
\be
V_M = \frac{A_M^L + A_M^R}{2}\,,\qquad
A_M = \frac{A_M^L - A_M^R}{2}\,,
\label{VAdefsmain}
\ee
appearing both in the singlet and nonsinglet flavor sectors that we describe in the following. We work in the
axial gauge with $V_r=0=A_r$. We can take the vector fluctuation to be transverse, $ \partial^\m  V_\mu = 0$, and
separate the axial vectors in transverse and longitudinal parts as
$A_\m (x^\mu, r)=A^\bot_\m (x^\mu, r)+ A^{\lVert}_\m(x^\mu, r)$
with $ \partial^\m A^\bot_\m = 0$.
We also write the complex tachyon field as
\be \label{Tfluctdef}
T(x^\mu,r)=\left[\tau(r)+s(x^\mu,r)+\mathfrak{s}^a(x^\mu,r)t^a\right]\,
\exp\left[i\theta(x^\mu,r)+i\,\pi^a(x^\mu,r)t^a\right]\,,
\ee
where $t^a$ are the generators of $SU(N_f)$, $\tau$ is the background solution, $s$ ($\theta$) is the scalar
(pseudoscalar) flavor singlet fluctuation, and $\mathfrak{s}^a$ ($\pi^a$) are the scalar (pseudoscalar) flavor
nonsinglet fluctuations.

\subsection{Nonsinglet fluctuations}

 The nonsinglet fluctuations include the vector and axial vector meson fluctuations 
(\ref{VAdefsmain}), the pseudoscalar mesons (including the massless pions), and the
scalar mesons.
Their second order equations are relatively simple, and we present those of the vectors below. We use the
standard factorized Ansatz $V_\mu (x^\mu, r) =  \psi^V(r)\, {\cal V}_\mu (x^\mu)$.
The radial wave function satisfies the following equation:
\be
{\partial_r \left( V_f(\l,\tau) \gf(\l)^2 e^{A}
G^{-1}\,  \partial_r \psi^V \right)\over {V_f(\l,\tau) \gf(\l)^2 e^{A} G}}
+m_V^2 \psi^V  = 0\,,\quad G \equiv \sqrt{1+ e^{-2A}\h(\l) (\partial_r \tau)^2 }\,.
\label{vectoreom}
\ee

 \begin{figure}[!tb]
\begin{center}
\includegraphics[width=0.49\textwidth]{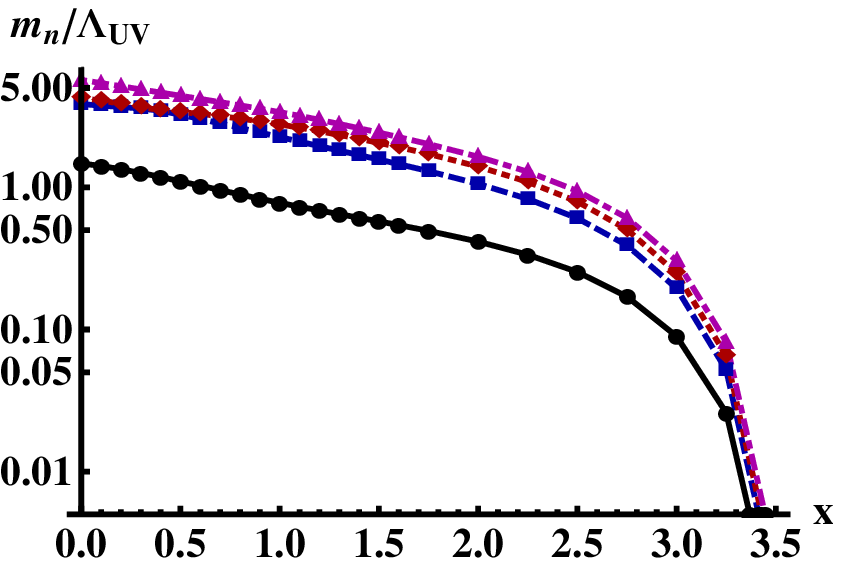}\hfill
\includegraphics[width=0.49\textwidth]{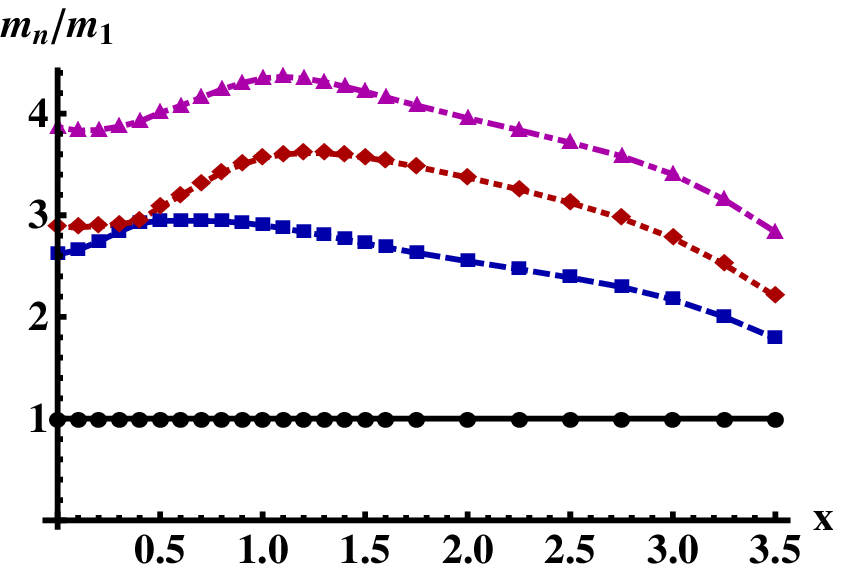}
\end{center}
\caption{\small Singlet scalar meson spectra for the potential II class with SB normalization for $W_0$.
They contain the $0^{++}$ glueballs and the singlet $0^{++}$ mesons that mix here at leading order.
Left: the four lowest masses as a function of $x$ in units
of $\Lambda_{\rm UV}$. Right: the ratios of masses of up to the fourth massive states as a function of $x$.
}
\label{f4}\end{figure}
 
The radial wave function for the transverse axial fluctuations can be defined by
$A^\bot_\m (x^\mu, r) = \psi^A(r)\, {\cal A}_\mu (x^\mu)\,,$ and it satisfies
\be
{\partial_r \left( V_f(\l,\tau) \gf(\l)^2 e^{ A}
G^{-1}\, \partial_r \psi^A\right)\over V_f(\l,\tau) \gf(\l)^2 e^{ A} G  }
-{4\tau^2 e^{2 A} \h(\l) \over \gf(\l)^2 }\psi^A+m_A^2 \psi^A = 0 \ .
\label{axvectoreom}
\ee
The nonsinglet scalar and pseudoscalar fluctuation equations are more complicated and we will present
them in \cite{to}. 

\begin{figure}[!tb]
\begin{center}
\includegraphics[width=0.49\textwidth]{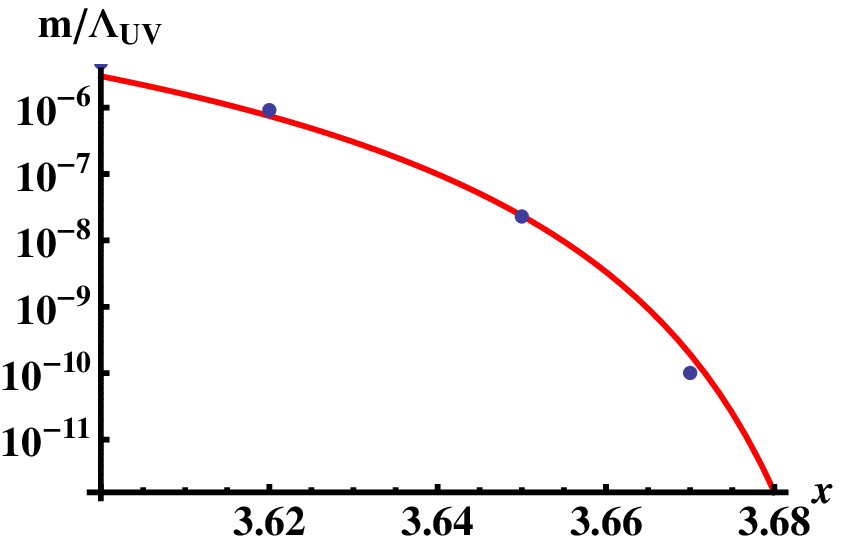}\hfill
\includegraphics[width=0.49\textwidth]{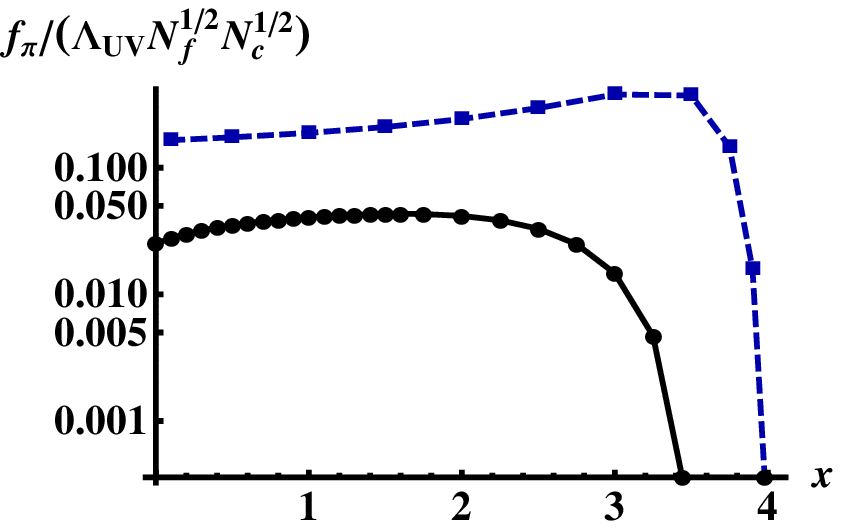}
\end{center}
\caption{\small Left: A fit of the $\rho$ mass to the Miransky
scaling factor for Potentials II with SB normalization for $W_0$. 
Right: $f_{\pi}$ as a function
of $x$ in units of $\Lambda_{\rm UV}\sqrt{N_cN_f}$. Again, it vanishes near $x_c$ following Miransky scaling.
The dashed blue curve is the result for potentials~I with $W_0=3/11$, while the continuous black curve is for
potentials~II with SB normalized $W_0$.}
\label{f5}\end{figure}

The general behavior of the spectra in the nonsinglet sector is as follows:
\begin{itemize}

\item In the conformal window all spectra are continuous.

\item Below the conformal window, $x<x_c$, the spectra are discrete and gapped.
The only exception being the $SU(N_f)$ pseudoscalar pions that are massless due to chiral symmetry breaking.

\end{itemize}

In the ``walking regime'', i.e., $x_c-x \ll 1$, we find the following specific features:
\begin{itemize}

\item All masses  obey Miransky scaling:
$m_n\sim \Lambda_{\rm UV} \exp({-{\kappa \over \sqrt{x_c-x}}})$. This is explicitly seen in the case
of the $\rho$ mass in figure \ref{f5} left.

\item All nonsinglet mass ratios asymptote to non-zero constants as $x\to x_c$.

\end{itemize}

We present the results of our numerical analysis of the nonsinglet meson spectra in figures \ref{f2} and \ref{f3}
(note that the plots on the left of those figures are in logarithmic scale). These results reflect the properties of
the spectra listed above.
The lowest masses of the mesons vary little with $x$ until we reach the walking region.
There, Miransky scaling takes over and the lowest masses dip down exponentially fast.
The $\Lambda_{UV}$ scale is extracted as usual from the logarithmic running of $\l$ in the UV.

\subsection{Singlet fluctuations}

The singlet fluctuations consist of the $2^{++}$ glueballs, the $0^{++}$ glueballs and scalar
mesons that mix to leading order in $1/N$ in the Veneziano limit, the $0^{-+}$ glueballs, and the
$\eta'$ pseudoscalar tower.
The spin-two fluctuation equations are simple and can be summarized by the appropriate
Laplacian (see for instance \cite{glue}).
The scalar and pseudoscalar equations are, however, very involved and
will be presented in detail in \cite{to}. Here we 
show results of the numerical analysis for the $0^{++}$ singlet scalars in figure~\ref{f4}.

The general properties of the singlet spectra are similar as in the nonsinglet sector: below the conformal
window, $x<x_c$, the singlet spectra are discrete and gapped, and there is again Miransky scaling as
$x \to x_c$ from below (see figure~\ref{f4}). 

There are some specific properties related to the mixing of the glueball and meson states:
\begin{itemize}

\item 
The $U(1)_A$ anomaly appears at leading order in the Veneziano limit, and consequently the mixture of the
$0^{-+}$ glueball and the $\eta'$ has a mass of ${\cal O}(1)$.

\item In the scalar sector, for small $x$, where the mixing between glueballs and mesons is small, the
lightest state is a meson, the next lightest state is a glueball, the next a meson and so on.
However, with increasing $x$, nontrivial mixing sets in and level-crossing seems to be generic.
This can be seen in the right hand plot of figure \ref{f4}.

\end{itemize}

\begin{figure*}[!tb]
\begin{center}
\includegraphics[width=0.40\textwidth]{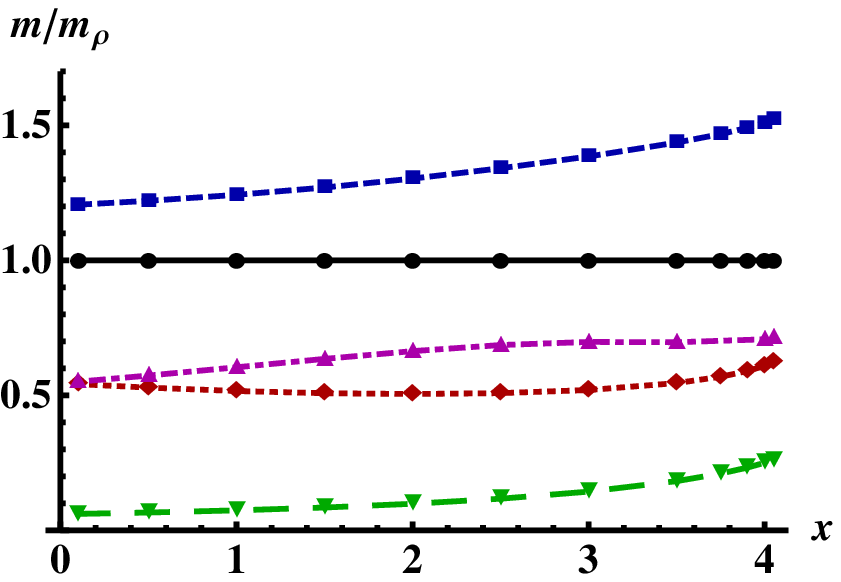}
\includegraphics[width=0.40\textwidth]{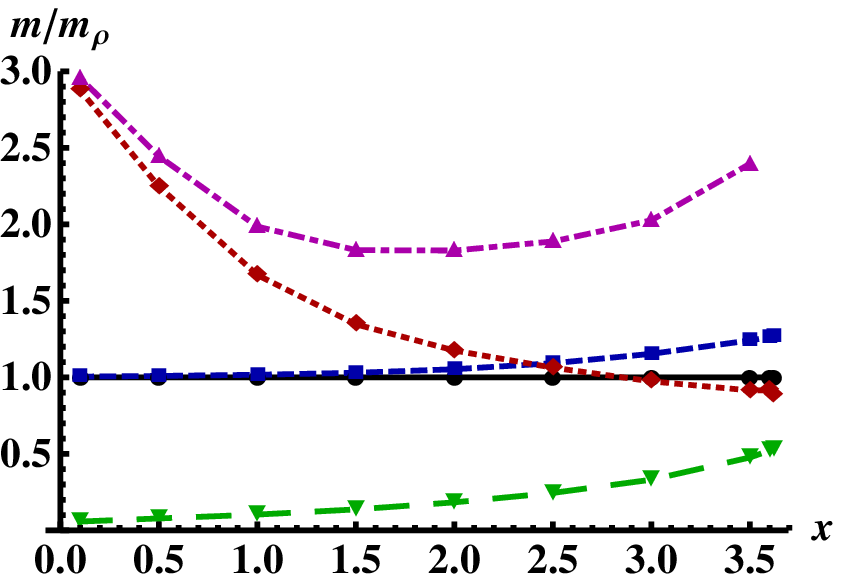}
\end{center}
\caption{\small  The masses of the lightest states of various towers, and $f_{\pi}/\sqrt{N_fN_c}$ as a function
of $x$ in units of the $\rho$ mass. Left: potentials I with $W_0=3/11$. Right: potentials II with SB normalization
for $W_0$. Solid black, dashed blue, dotted red,  and dotdashed magenta curves show the masses of the
lightest vector, axial, flavor nonsinglet scalar, and flavor singlet scalar states, respectively, while the long-dashed
green curve is $f_{\pi}/\sqrt{N_fN_c}$ (see section~\protect\ref{sec:Sfpi}).}
\label{fratios}\end{figure*}

All singlet mass ratios asymptote to constants as $x\to x_c$. The same holds for mass ratios between the
flavor singlet and nonsinglet sectors, as confirmed numerically in figure~\ref{fratios}.
There seems to be no unusually light state (termed the ``dilaton'') that reflects the nearly broken scale
invariance in the walking region. The reason is a posteriori simple: the nearly broken scale invariance is
reflected in the {\em whole} spectrum of bound states scaling exponentially to zero due to Miransky scaling.

\subsection{Asymptotics of the spectra}

The asymptotics of the spectra at high masses is in general a power-law with logarithmic corrections, with the
powers depending on the potentials.
The trajectories are approximately linear ($m_n^2\sim c n$) for type I potentials
  and quadratic ($m_n^2\sim c n^2$) for type II potentials.
 There is the possibility, first seen in \cite{ckp} that the proportionality coefficient $c$ in the linear case is different
 between axial and vector mesons\footnote{A careful analysis of the effects of different potentials
 on the asymptotics  of the spectra will be presented in \cite{to}.}.  These possibilities do not affect substantially
 the issues of the dilaton and the S-parameter. 
 
Finally, let us comment on the possibility of using as background the nontrivial saddle points, found in \cite{jk},
where the tachyon solution has at least one zero (analogous to the Efimov minima). We have verified explicitly that
such saddle points are unstable, as the scalar meson equation has a single mode with a negative mass squared,
both in the singlet and nonsinglet channels. This mass is small for small $x$, but becomes large as
$x\to x_c$ \cite{kutasov}. Therefore, the Efimov minima are strongly unstable in the walking regime.

\begin{figure}[!tb]
\begin{center}
\includegraphics[width=0.49\textwidth]{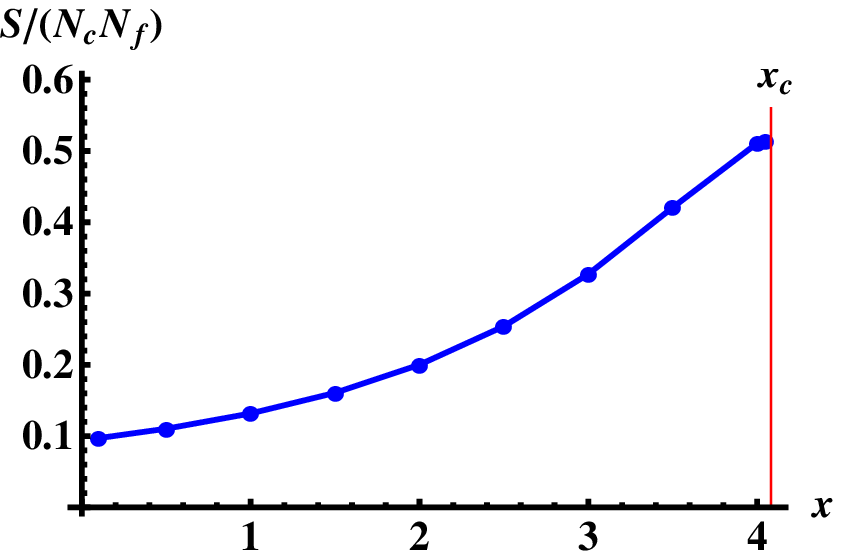}\hfill
\includegraphics[width=0.49\textwidth]{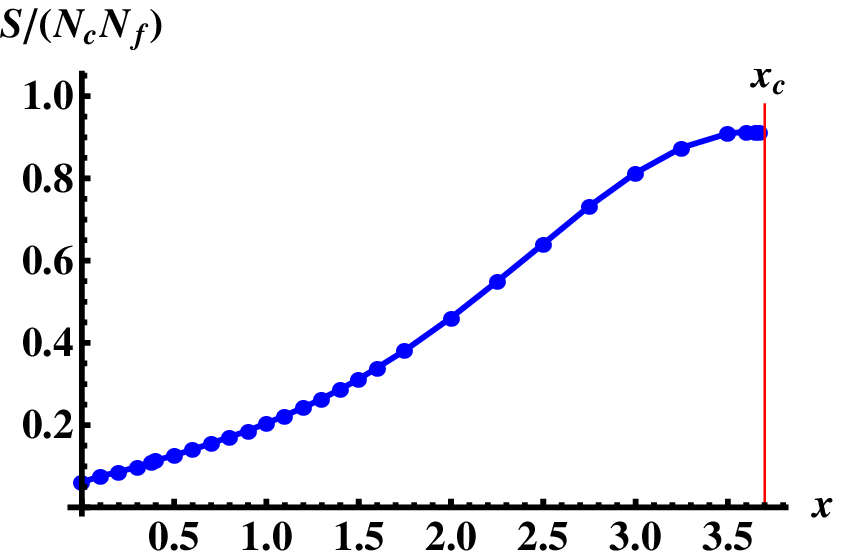}
\end{center}
\caption{\small  Left: The S-parameter as a function of $x$ for potential class I with $W_0={3/ 11}$.
  Right: The S-parameter as a function of $x$ for potential class II with SB normalization for $W_0$.
In both cases $S$ asymptotes to a finite value as $x\to x_c$.
}
\label{f6}
\end{figure}

\section{Two-point functions and the S-parameter} \label{sec:Sfpi}

We have computed the two-point functions of several operators including the axial and vector currents  as well
as the scalar mass operator. We will focus here on the two-point functions of the vector and axial currents which
can be written in momentum space as
\be
\langle V_{\m}^{a}(q) V_{\n}^{b}(p) \rangle= \Pi^{ab}_{\mu \nu,V}(q,p)=-(2 \pi)^4 \delta^4 (p+q)\,
\delta^{ab} \left( q^2\eta_{\m\n}-{q_{\m} q_{\n}}\right)\Pi_{V} (q) \,,
\ee
and similarly for the axial vector. Here we have used that
\be
V_{\m}(x)=\int {d^4 q \over (2\pi)^4}\, e^{i q x}\, V_{\m}^{a}(q)\, t^a \, \psi_V(r)\,,
\ee
where $t^a$, $a=1,\ldots, N_f^2-1$ are the flavor group generators.

Next, using the expansions
\be
\Pi_A = {f_{\pi}^2 \over q^2} +\sum_{n} {f_n^2 \over q^2 +m_n^2 - i \epsilon}\,,\qquad
\Pi_V=\sum_{n}{F_n^2 \over q^2 +M_n^2 - i \epsilon}\,,
\label{2}\ee
we determine $f_{\pi}$ as
\be
f_{\pi}^2= -{N_{c} N_{f} \over 12 \pi^2 } \left. {\partial_{r} \psi^A
    \over r} \right|_{r=0,\, q=0}\,,
\ee
where the normalization is fixed by matching the UV limit of the two point functions to QCD. Here the normalization
of the radial wave function was fixed in the UV by $\psi^A(r=0)=1$, and it is required to be normalizable in the IR.

Typical results for $f_{\pi}$ are plotted in figure \ref{f5}, right.
Notice that the pion scale changes smoothly for most of the range of $x$, but as $x\to x_c$ Miransky scaling sets
in such that it vanishes exponentially.

Finally, we shall now compute the S-parameter for V-QCD. It is given by:
\bea
S=4 \pi {d \over dq^2}\left[q^2 (\Pi_V - \Pi_A)\right]_{q=0}& =&-{N_c N_f \over 3 \pi} {d \over dq^2}\left.
\left( {\partial_r \psi^V (r) \over  r}-{\partial_r \psi^A (r) \over r} \right) \right|_{r=0,\, q=0}
\nonumber\\
&=&4\pi\sum_{n}\left({F_n^2\over M_n^2}-{f_n^2\over m_n^2}\right) \ .
\label{3}
\eea

As both masses and decay constants in
(\ref{3}) obey Miransky scaling, the S-parameter is insensitive to it.
Therefore subleading terms determine its scaling behavior as $x \to x_c$. 
Our results show that generically the S-parameter (in units of $N_fN_c$) remains finite in the QCD regime,
$0<x<x_c$, and asymptotes to a finite constant at
$x_c$ (see figure~\ref{f6}). The S-parameter is identically zero inside the conformal window (massless quarks)
because of unbroken chiral symmetry. This suggests a subtle discontinuity of correlators across the conformal
transition, which will be analyzed in detail in \cite{to}. In \cite{parnachev}
similar conclusions are reached in a different context (probe tachyon-flavor dynamics in
AdS). We find that in the walking region of V-QCD the backreaction of flavor to matter (that is fully implemented here)
is important,  among other things,  for the spectra, and therefore the two results are not directly comparable.

This behavior of $S$ is in qualitative agreement with recent estimates based on analysis of the BZ limit in field
theory \cite{sannino}. We have also found choices of potentials
where the S-parameter becomes very large as we approach $x_c$.
Our most important result is that generically the S-parameter is an increasing function of $x$, and reaches its
highest value at or near $x_c$ contrary to previous expectations \cite{sannino}.

\section{Conclusions}

In this letter we have studied the zero temperature spectra of glueballs and mesons in a class of holographic
theories (V-QCD) that is in the universality class of QCD in the Veneziano limit. This model takes into account the
backreaction of the flavor degrees of freedom and therefore allows us to analyze the spectra as a function of
$x=N_f/N_c$. 

V-QCD was formulated in \cite{jk} where the zero temperature phase diagram was studied and
shown to agree with expectations for QCD in the Veneziano limit. 
It is for these solutions that we have computed the spectra of fluctuations. We have found that the main features
of the spectra are shared by various choices of the potentials, although some important issues like the large mass
asymptotics do depend on their specific details. 
A thorough study of the
spectra for a wider class of potentials is underway and we expect to report its results in the near future \cite{to}.
The study presented here has nevertheless allowed us to outline the main characteristics of the spectra of
fluctuations and also to address issues relevant for the use of this setup as a holographic model
of walking technicolor. We will now summarize our main results.

In the conformal window, $x_c<x<11/2$ where the theory flows to an IR nontrivial fixed point, all the spectra are
continuous. For $x<x_c$ the massless theory spontaneously breaks chiral symmetry in the IR. The spectra in
this region are discrete and gapped (except for the pions). For $x\lesssim x_c$ the model is in its ``walking
region'', where the coupling stays almost constant for many decades of energy and it only diverges in the deep
IR. As for the spectra, in the ``walking region'' all masses obey Miransky scaling:
$m_n\sim \Lambda_{\rm UV} \exp({-{\kappa \over \sqrt{x_c-x}}})$, and the same applies to other mass
parameters like $f_\pi$. 

Remarkably, as $x\to x_c$ all flavor singlet and nonsinglet mass ratios asymptote to
non-zero constants. Therefore, there is no unusually light state (also called ``dilaton'') in the spectrum.
Such a state was expected as a consequence of the approximate conformal symmetry in the ``walking region''.
In this scenario this approximate conformal symmetry is instead correlated with Miransky scaling of all masses.
One should also notice that for finite values of $x$ there is strong mixing between singlet mesons and glueballs,
and occasional level crossing as $x$ is varied.

Finally, by computing the two-point functions of vector and axial vector mesons, we have been able to determine
the S-parameter for our setup. In units of $N_f\,N_c$ the S-parameter is generically of ${\cal O}(1)$. Additionally, it 
is an increasing function of $x$ and asymptotes to a finite constant as $x\to x_c$. Inside the conformal window the
S-parameter is identically zero and therefore it is discontinuous at 
the conformal transition (at $x=x_c$).

These results for the S-parameter suggest that making $S$ arbitrarily small in a walking theory may be more
difficult than expected before. Moreover, our results indicate that this is probably not the case for QCD in the
Veneziano limit.

 \addcontentsline{toc}{section}{Acknowledgments}
\acknowledgments
We thank the organizers of the ``XVIII European Workshop on String Theory'' held at the Corfu Summer Institute
for the oportunity to present our work there.

This work was in part supported by grants
 PERG07-GA-2010-268246, PIEF-GA-2011-300984, the EU program ``Thales'' ESF/NSRF 2007-2013, and by the European Science
Foundation ``Holograv'' (Holographic methods for strongly coupled systems) network.
It has also been co-financed by the European Union (European Social Fund, ESF) and
Greek national funds through the
 Operational Program ``Education and Lifelong Learning'' of the National Strategic
 Reference Framework (NSRF) under
 ``Funding of proposals that have received a positive evaluation in the 3rd and 4th Call of ERC Grant Schemes''.
D.A. would like to thank the Crete Center for Theoretical Physics for hospitality and the FRont Of
pro-Galician Scientists for unconditional support. I. Iatrakis' work was supported by the project
``HERAKLEITOS II - University of Crete'' of the Operational Programme for Education and Lifelong Learning
2007 - 2013 (E.P.E.D.V.M.) of the NSRF (2007 - 2013), which is co-funded by the European Union
(European Social Fund) and National Resources.


\end{document}